\documentclass[aps,pra,numerical,reprint,showpacs]{article}

\usepackage{graphicx}
\usepackage{bm}
\usepackage{amsmath}
\usepackage{amssymb}
\usepackage[colorlinks=true, linkcolor=blue, citecolor=red, urlcolor=blue]{hyperref}

\makeindex             

\begin{document}

\title{Parametrisation in electrostatic DPD Dynamics and Applications}

\author{E. Mayoral$^\dagger$ and E. Nahmad-Achar$^\ddagger$}

\maketitle
$^\dagger$
Instituto Nacional de Investigaciones Nucleares, Carretera M\'exico-Toluca S/N, La Marquesa Ocoyoacac, Edo. de M\'exico
C.P. 52750, M\'exico\\
$^\ddagger$
Instituto de Ciencias Nucleares, Universidad Nacional Aut\'onoma de M\'exico, Apartado Postal 70-543, 04510 M\'exico DF,   Mexico  \\


\section*{abstract}
A brief overview of mesoscopic modelling via dissipative particle dynamics is presented, with emphasis on the appropriate parametrisation and how to calculate the relevant parameters for given realistic systems. The dependence on concentration and temperature of the interaction parameters is also considered, as well as some applications.

\section{Introduction}

In a colloidal dispersion, the stability is governed by the balance
between Van der Waals attractive forces and electrostatic repulsive forces, together with steric mechanisms. Being able to model their interplay is
of utmost importance to predict the conditions for colloidal
stability, which in turn is of major interest in basic research and
for industrial applications.

Complex fluids are composed typically at least of one or more solvents, polymeric or non-polymeric surfactants, and crystalline substrates onto which these surfactants adsorb. Neutral polymer adsorption has been
extensively studied using mean-field approximations and assuming an adsorbed polymer configuration of loops and tails~\cite{deGennes, deGennes2, deGennes3, mendez}. Different mechanisms of adsorption affecting the global stability of a colloidal dispersion, including surface-modifying polymer chains {\it versus} end-grafted polymer chains, have been studied in~\cite{langmuir}. Attempts to measure the forces themselves that act in a confined complex fluid in thermodynamic equilibrium with its surroundings have been made using atomic force microscopy (cf., e.g.,~\cite{McNamee}), while it has been argued~\cite{derjaguin} that it is more appropriate to use the concept of disjoining pressure, which is the difference between the force (per colloidal particle unit area) normal to the conﬁning surfaces and the
fluid’s bulk pressure. This disjoining pressure allows for a direct determination of the free energy of interaction, hence its importance.

Polyelectrolyte solutions have very different properties from those observed
in solutions of uncharged polymers, and their behaviour is less well known~\cite{deGennes4, odjik, dobrynin, mayoral1}. In particular, it is not evident that the scaling of some quantities present a similar behaviour as that of electrically neutral solutions, or that they present the same or similar scaling exponents. Calculating Langmuir isotherms for polyacrylate dispersants adsorbed on metallic oxides, and their scaling properties as a function of the number of monomeric dispersant units obtained via dissipative particle dynamycs (DPD) simulations, it has been shown~\cite{mayoralSpringer2} that the critical exponent for the renormalized isotherms agrees perfectly well with the scaling theory in~\cite{deGennes5} even though polyelectrolytes were being considered.

Due to the long-range Coulombic repulsion produced by the presence
of small mobile counterions in the bulk, the properties of these systems cannot in general be obtained analytically. The most usual systems are even more complex, encompassing various surfactants of different chemical nature and molecular weight (acting as dispersants, wetting agents, rheology modifiers,e tc.), pigments, ``inert'' extenders, and so on. In all these cases there are various different length and dynamic scales, every species interact with all others at a molecular level, in a way which is dependent on temperature and concentration. There is, further, competitive adsorption amongst all surfactants present. Ideally, one should have a basic understanding of all interactions, but the main problem is that all colloidal systems are thermodynamically unstable. Empirical methods have been used as well as few and greatly approximated analytic models, and a more recent and promising method is that of {\it molecular dynamics simulations}. Its basic methodology consists of taking advantage of the fast computing facilities that are nowadays available, to integrate Newtons equations of motion for a large number $N$ of particle (molecules, atoms, or whatever the problem in turn calls for). Thus, one sets initial positions $r_i(t)$ and momenta $p_i(t)$ for each particle $i$ at time $t$, and uses the force field felt by each one of them
\begin{equation}
    F(r) = - \nabla\,V(r) = m\,\frac{d^2r}{dt^2}
    \label{Newton}
\end{equation}
to find its new position and momentum at time $t+\delta t$ iteratively. The approximation being made is to consider the potential $V(r)$ to be constant during the time step $\delta t$ which, if taken very small, can make the error negligible. Typical choices for the force field are the electrostatic interaction $V(r) = k\,q\,q'/r$ and a Lennard-Jones type potential $V(r) = 4\varepsilon\,\left[(\sigma/r)^{12} - (\sigma/r)^{6}\right]$, where the adjustable parameters $(\varepsilon,\ \sigma,\ k)$ must be obtained by other means (first principles or experimentation). Relevant quantities of the system are computed as time-averages over a macroscopic time interval
\begin{equation}
    A = \lim_{t\to\infty}\, \frac{1}{t}\,\int_{t_0}^{t_0+t}\, A\left[r_1(t'), r_2(t'),...,r_N(t');\, p_1(t'), p_2(t'),...,p_N(t')\right]\, dt'\ .
\end{equation}

\begin{figure}
\begin{center}
\scalebox{0.4}{\includegraphics{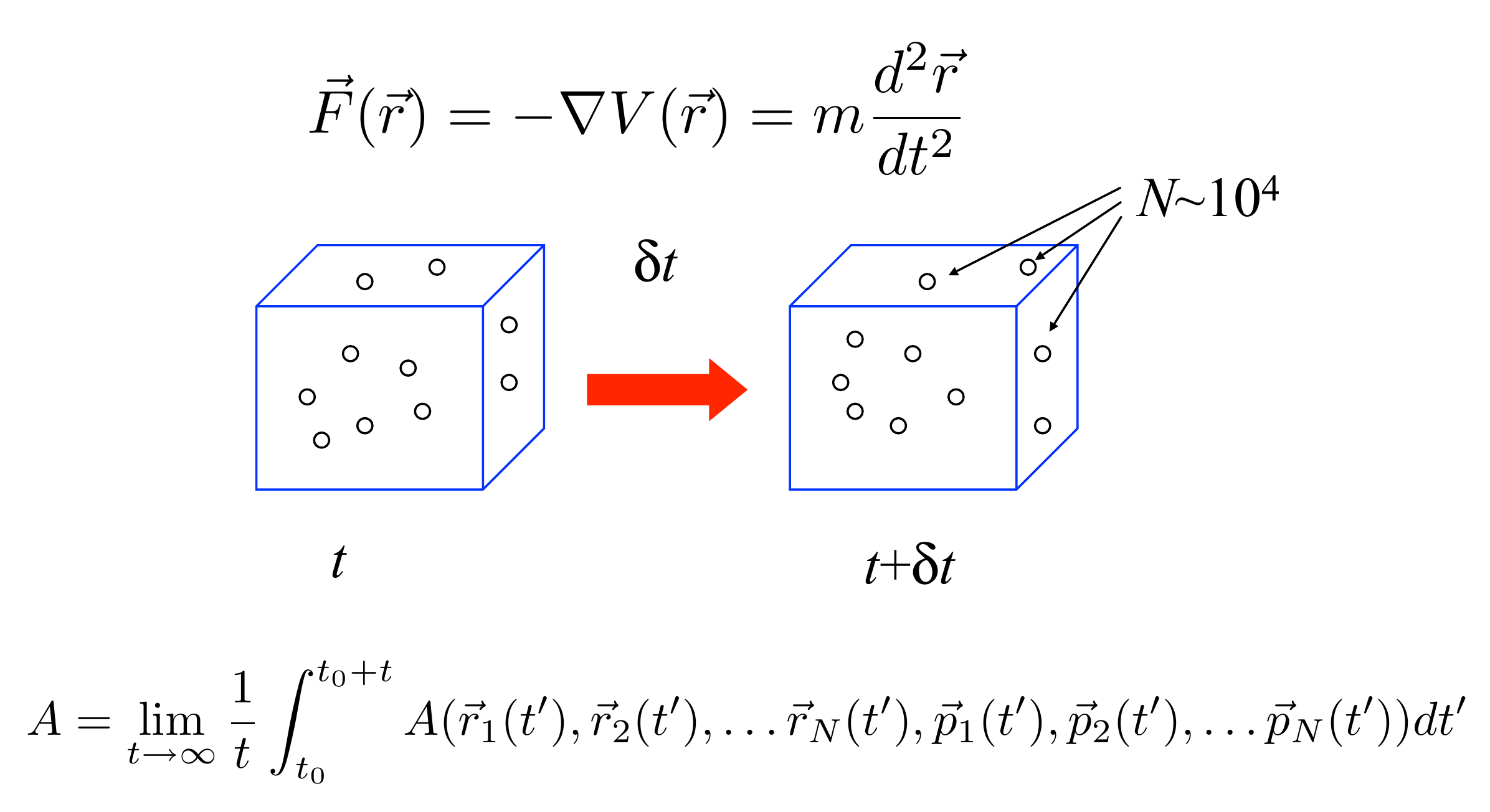}}
\end{center}
\caption{Descriptive methodology of a {\it molecular dynamics simulation} (see text for details).}
\label{fig1}
\end{figure}

The pieces of information that one can obtain through these simulations are mainly structural and thermodynamic properties:  {\it i)} the density profile $\rho(r)$, which in particular may be used to characterise when two phases (e.g. liquid and vapour) coexist; {\it ii)} the radial distribution function $g(r)$ given by
\begin{equation}
    \rho(r) = \int\,\langle\rho\rangle\,g(r)\,dr
\end{equation}
which measures the average number of particles in each coordination shell with respect to a given centre (and usually obtained through X-ray or neutron scattering experiments); {\it iii)} the interfacial tension
\begin{equation}
    \gamma^\ast = L_z\,\left[P_{zz}-\frac{1}{2}\left(P_{xx}+P_{yy}\right)\right]
    \label{interfacial}
\end{equation}
obtained from the pressure tensor components $P_{ij}$ within a box of length $L_z$; {\it iv)} the radius of gyration of a polymer chain, given by
\begin{equation}
    R_g = a_f\,\mathcal{N}^\nu
\label{rgscaling}
\end{equation}
where $a_f{}^3$ is proportional to the Flory volume, $\mathcal{N}$ is the monomer length of the chain, and $\nu$ is the appropriate scaling exponent;
{\it v)} phase diagrams; {\it vi)} adsorption isotherms; {\it vii)} disjoining pressures; etc. Figure~\ref{fig1} shows descriptively this methodology.

By its nature, microscopic molecular dynamics simulations require a great deal of computational resources, the reason being that the integration of the equations of motion require very small time steps when the interaction potential changes significantly over small distances. An alternative that has proven to be very successful is to do mesoscopic modelling via {\it dissipative particle dynamics} (DPD)~\cite{hoogerbrugge}, consisting of carrying out a coarse-graining of the microscopic degrees of freedom. It is highly dependent on parameters describing the different kinds of force fields, the parametrisation of which are not always clear in the literature. For this reason, we present here a revision of DPD parametrisation together with applications and comparison with experimental results.

In Section 2 we give a brief description of the DPD modelling, including {\it electrostatic} DPD. Section 3 deals with the appropriate parametrisation and how to calculate the relevant parameters for given realistic systems. The dependence on concentration and temperature of the interaction parameters is also considered. Section 4 presents some interesting applications, and we close with some Conclusions.

\section{Electrostatic Dissipative Particle Dynamics: a brief overview}

A good alternative to overcome the difficulties presented by molecular dynamics simulations is to do a coarse-graining of the microscopic degrees of freedom. When done carefully, results can be obtained which approximate very well those obtained through lengthy experimentation (cf. e.g.~\cite{mayoral1,langmuir,mayoral2, mayoral3} and references therein). The method of dissipative particle dynamics (DPD), introduced by Hoogerbrugge and Koelman~\cite{hoogerbrugge}, consists of grouping several molecules, or parts of molecules, in a representative way, into soft mesoscopic ``particles''. As with molecular dynamics simulations, one integrates the equations of motion to obtain the particle's positions and velocities, but here one distinguishes only between $3$ contributions to the total force: {\it conservative}, {\it dissipative} and {\it random}.

Conservative forces account for local hydrostatic pressure and are of the form
\begin{equation}
    \bm{F}_{ij}^c =
    \begin{cases}
    a_{ij}\,\omega^c(r_{ij})\,\hat{\bm{e}}_{ij}, &\text{$(r_{ij} < r_c)$} \\
    0, &\text{$(r_{ij} \geq r_c)$}.
     \end{cases}
\end{equation}
Here, $a_{ij}$ is a parameter which represents the maximum repulsion between particles $i$ and $j$, $\bm{r}_{ij} = \bm{r}_i-\bm{r}_j$, $r_{ij} = \vert \bm{ r}_{ij} \vert$, and $\hat{\bm{e}}_{ij} = \bm{r}_{ij}/r_{ij}$ where $\bm{r}_i$ denotes the position of particle $i$, and the weight function is given by $\omega^c(r_{ij}) = 1-r_{ij}/r_c$.

\begin{figure}
\begin{center}
\scalebox{0.6}{\includegraphics{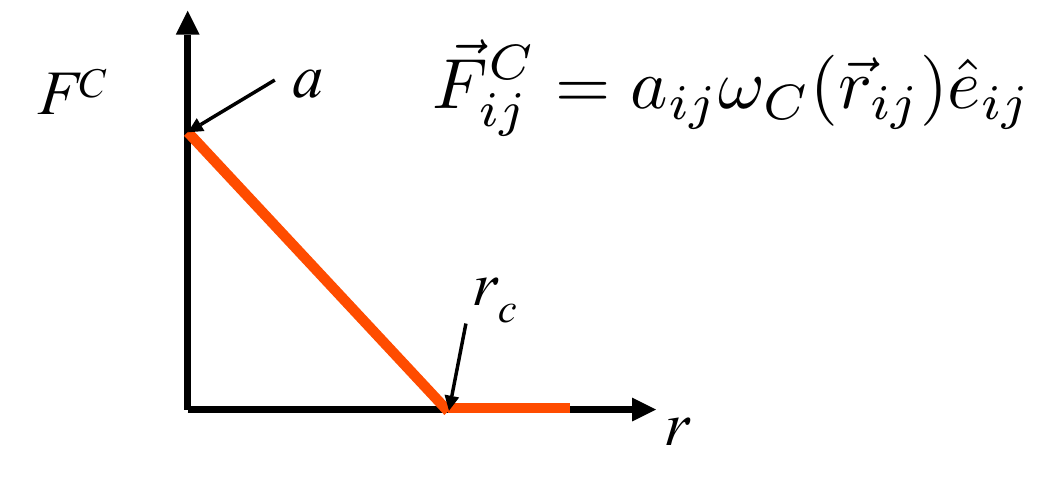}}
\end{center}
\caption{Form of the conservative force in the DPD methodology.}
\label{fig2}
\end{figure}

This force, depicted in Figure~\ref{fig2}, derives from a soft interaction potential and there is no hard-core divergence as in the case of the Lennard-Jones potential, which makes more efficient the scheme of integration since it allows for a large time step. In the case of macromolecules, such as polymers, the particles (which can consist of representative monomers or sets of monomers) are joined by springs with a spring constant $k$, so we have an extra conservative force of the form $\bm{f}_{ij} = -k\,\bm{r}_{ij}$ whenever particle $i$ is connected to particle $j$.

Dissipative forces account for the local viscosity of the medium, and are of the form
\begin{equation}
    \bm{F}^D_{ij} = -\gamma\,\omega^D(r_{ij})\,\left[ \hat{\bm{e}}_{ij} \cdot \bm{v}_{ij} \right]\,\hat{\bm{e}}_{ij}
\end{equation}
where $\bm{v}_{ij}=\bm{v}_i-\bm{v}_j$ is the relative velocity, $\gamma$ the dissipation constant, and $\omega^D(r_{ij})$ a dimensionless weight function.

Finally, the random (thermal) force disperses heat produced by the dissipative force and invests it into Brownian motion in order to keep the temperature $T$ locally constant. It is of the form
\begin{equation}
    \bm{F}^R_{ij} = -\sigma\,\omega^R(r_{ij})\,\xi_{ij}\,\hat{\bm{e}}_{ij}
\end{equation}
with $\xi_{ij} = \theta_{ij}\,(1/\sqrt{\delta_t})$, where $\delta_t$ is the integration time-step and $\theta_{ij}$ is a random Gaussian number with zero mean and unit variance. A dimensionless weight function $\omega^R(r_{ij})$ also appears.

Not all three forces are independent. The fact that the random force compensates the energy dissipated in order to keep $T$ constant means that it acts as a regulating thermostat. This leads to the {\it fluctuation-dissipation} theorem~\cite{warren} which gives
\begin{equation}
    \gamma = \frac{\sigma^2}{2\,k_B\,T}, \quad \omega^D(r_{ij}) = \left[\omega^R(r_{ij})\right]^2
\end{equation}
where $k_B$ Boltzmann’s constant.

When dealing with electrically charged species, such as polyelectrolytes, a problem with the DPD methodology, arising from the fact that the interactions are soft, is the artificial formation of ionic clusters. Electric charges are usually treated as point charges whose potential diverges at their position in space. In~\cite{mayoral1} this problem is solved by considering charge distributions over the DPD-particles. Suppose that we have a system constituted by N particles, each one with a point charge $q_i$ and a position $r_i$ in a volume $V = L_x\, L_y\, L_z$. Charges interact according to Coulomb's law and the total electrostatic energy for the periodic system is given by
\begin{equation}
    U(r^N) = \frac{1}{4\pi\varepsilon_0\varepsilon_r} \left[ \sum_i \sum_{j>i} \sum_{n_x} \sum_{n_y} \sum_{n_z} \frac{q_i\,q_j}{\vert \bm{r}_{ij} + (n_xL_x,\,n_yL_y,\,n_zL_z)\vert} \right]
\end{equation}
where $\bm{n} = (n_x,\, n_y,\, n_z)$, $n_x,\ n_y,\ \text{and}\ n_z$ are non-negative integer numbers, and $\varepsilon_0\ \text{and}\ \varepsilon_r$ are the dielectric constants of vacuum and water at room temperature, respectively. It is convenient to decompose this long-range electrostatic interaction into real and reciprocal space, getting a short-ranged sum which may be written as
\begin{eqnarray}
    U(r^N) = \frac{1}{4\pi\varepsilon_0\varepsilon_r} \Big[ && \sum_i \sum_{j>i} q_i\,q_j \frac{\text{erfc}(\alpha_\varepsilon r)}{r} + \nonumber \\
&& + \frac{2\pi}{V}\,\sum_{k\neq 0}^\infty Q(k)\,S(k)\,S(-k) - \frac{\alpha_\varepsilon}{\sqrt{\pi}}\sum_{i=1}^N q_i{}^2 \Big]
\end{eqnarray}
with
\begin{equation}
    Q(k) = \frac{e^{-k^2 / 4\alpha_\varepsilon^2}}{k^2}, \quad S(k) = \sum_{i=1}^N q_i\,e^{i\,\bm{k}\cdot r_{ij}}, \quad
    \bm{k} = \frac{2\pi}{L}\,(m_x,\, m_y,\, m_z) \nonumber
\end{equation}
Here, $\alpha_\varepsilon$ is the parameter that controls the contribution of the real space, $k$ is the magnitude of the reciprocal vector $\bm{k}$, $m_x,\ m_y,\ m_z$ are integer numbers, and $\text{erfc}(\alpha_\varepsilon r)$ is the complementary error function (cf.~\cite{mayoral2}).

The various parameters introduced, viz. $a_{ij},\ \sigma,\ \gamma,\ \theta_{ij}$, contain all the information of the particular system being considered. It is therefore crucial, for the DPD methodology to work, to be able to establish these parameters faithfully.

\section{Parametrisation for Realistic Systems}
\label{parametrisation}

By far the most important parameter is the one defining the conservative force, $a_{ij}$, because it contains all the physicochemical information for each component in the system. In contrast, the noise and dissipative parameters correspond to the temperature and fluid viscosity respectively. In a mono-component system the conservative force parameter for equal species $a_{AA} \equiv a$ relates to the inverse isothermal compressibility~\cite{groot}
\begin{equation}
    \kappa^{-1} = \frac{1}{n\,k_B\,T\,\kappa_T} = \frac{1}{k_B\,T}(\partial p/\partial n)_T
\end{equation}
where $n$ is the number density of molecules and $\kappa_T = (\partial p/\partial n)_T$ is the usual isothermal compressibility. The pressure $p$ in the system may be obtained using the viral theorem, obtaining $p = \rho\,k_BT + \alpha\,a\,\rho^2$, where $\rho$ is the density and $\alpha = 0.101$ for $\rho > 2$. We then have $\kappa^{-1} = 1 + 2\alpha\,a\,\rho/k_BT \simeq 1 + 0.2a\,\rho/k_BT$. If $N_m$ is the number of molecules contained in a DPD particle, then $a = k_BT(\kappa^{-1} N_m - 1)/2\alpha\,\rho_{DPD}$, where $\rho_{DPD}$ is the DPD number density for the system and is usually set to  $\rho_{DPD}=3$ (three water molecules per mesoscopic particle in an aqueous solution, for example). For the mono-component system the virial free energy density $f_v$ is given by $f_v / k_BT = \rho \ln\rho - \rho + 2\alpha\,a\,\rho^2/k_BT$.

When a mixture of $2$ components $A$ and $B$ is considered, the virial pressure is given by~\cite{maiti}
\begin{equation}
    p = \frac{\alpha\,k_BT\,\rho^2}{r_c{}^3}\,\left[ a_{AA}\,\phi^2 + 2\,a_{AB}\,\phi(1-\phi) + a_{BB}(1-\phi)^2 \right]
    \label{virialpressure}
\end{equation}
where $\phi$ is the volume fraction of component $A$ and $(1-\phi)$ that of component $B$, and the virial free energy density for this system is
\begin{equation}
    f_v / \rho\,k_BT = \frac{\phi}{N_A}\,\ln\phi + \frac{(1-\phi)}{N_B}\,\ln(1-\phi) + \frac{\alpha(2\,a_{AB} - a_{AA} - a_{BB})\rho}{k_BT} \phi(1-\phi) + cte
    \label{virialfreeenergy}
\end{equation}
with $\rho = \rho_A + \rho_B$ and $a_{AB} = a_{BA}$.

The relationship between $a_{ij}$ and the physicochemical characteristics of a real system may be obtained through the Flory-Huggins (FH) theory, based on occupations of a lattice where we have exclusively and uniquely a polymer segment or a solvent molecule per lattice site. In the mean-field approximation this exacting single occupancy is relaxed to a site occupancy probability, which gives a mean-field free energy of mixing constituted by a combinatorial entropy and a mean-field energy of mixing $\Delta F^{MF}_{MIX} = \Delta S^{MF}_{MIX} + \Delta H^{MF}_{MIX}$. The free energy per unit volume for a mixture of two polymers $A$ and $B$ could then be written as
\begin{equation}
    \frac{\Delta F^{MF}_{MIX}}{N\,k_BT} = \frac{\phi}{N_A}\ln\phi + \frac{(1-\phi)}{N_B}\ln(1-\phi) + \chi(\phi)(1-\phi)
    \label{DeltaF}
\end{equation}
with $N_A$ and $N_B$ the number of monomers of species $A$ and $B$ respectively, and $N=N_A+N_B$. The first two terms on the right hand side contain the information of the energy of the pure components and correspond to the entropic contribution $\Delta S^{MF}_{MIX}$. The third one involves the excess energy produced by the mixture ($\Delta H^{MF}_{MIX}$). The $\chi$-parameter tells us how alike the two phases are, and is known as the {\it Flory-Huggins interaction parameter}. In the mean-field theory this parameter is written in terms of the nearest-neighbor interaction energies $\epsilon_{ij}$ as $\chi_{12}=z(\epsilon_{11} + \epsilon_{22} - \epsilon_{12})/2k_B T$, where $z$ is the lattice coordination number. It is a phenomenological parameter, and corrections considering an ionisation equilibrium between counterions and electrolyte are needed in the presence of long-range forces. But one can also estimate this quantity by using the Hildebrand-Scatchard regular solution theory~\cite{hildebrand, scatchard, hildebrand2}, in which the entropy of mixing is given by an ideal expression but the enthalpy of mixing is non-zero and is the next simplest approximation to the ideal solution. In this approach one can appropriately consider the Coulombic contribution in the enthalpy of mixing via the activity coefficients in electrolyte solutions ({\it vide infra}).

Whereas the FH mean-field theory considers $\chi_{12}$ as proportional to $T^{-1}$ but independent of the solute concentration $\zeta$, comparisons with experiments show that phenomenological $\chi_{12}$ contains both energetic and entropic contributions; i.e., $\chi_{12} = \chi_{12}(T,\,\zeta)$. A correct parametrisation in our electrostatic DPD system must, therefore, take into account the dependence of the repulsive parameters for the solvated ions $a_{ij}$ with the salt concentration $\zeta$. The way to understand this is as follows: when we perform a coarse graining, the volume of a DPD particle does not usually encompass a full molecule or polymer; thus, for instance, although for dodecane our DPD particle contains only a butane fragment, one does not construct dodecane from the union of butane particles, and the interaction between the DPD dodecane particles and water does not correspond with the $\chi$ parameter of butane with water; the $\chi$ parameter employed to estimate the DPD repulsive parameter $a_{ij}$ should be that of the full dodecane molecule because its behaviour is that of the global joined units which affect the electronic distribution throughout. In this case, the ``monomeric'' units which constitute the dodecane ``polymeric'' molecule interact through short-range (covalent bond) forces. When considering a solvated electrolyte, e.g. $N_a{}^{+}$ or $Cl^{-}$ ions, their concentration is given precisely by the amount of solvated ionic particles present, which corresponds effectively with the amount of ``monomeric'' solvated ionic units. These are in effect the individual DPD units, which in this case are not covalently joined but are subject to long-range electrostatic forces. The presence and quantity of ``monomeric'' solvated ions affect the global properties of the network and their corresponding $\chi$ parameter should take into account the whole electrolytic entity, and thus a correct parametrisation of the DPD system forces a dependence of the conservative force parameters $a_{ij}$ on the concentration $\zeta$, through $\chi(T,\,\zeta)$.

\subsection{Concentration dependence of the DPD interaction parameters}

For an electrolyte solution in water, $e\,+\,w$, the chemical potential $\mu_{w/e}$ for each component ($w/e$) may be obtained by differentiating the free energy per unit volume of the mixture $e\,+\,w$ with respect to the number of molecules $N_{w/e}$ of the component in question. Thus,
\begin{equation}
    \frac{\mu_w}{k_BT} = \ln\phi + \chi(1-\phi)^2, \quad \frac{\mu_e}{k_BT} = \ln(1-\phi) + \chi\phi^2
\end{equation}
where $\phi$ and $1-\phi$ are the volumetric fractions for the $w$ (solvent) and  $e$ (electrolyte) components respectively. The activity coefficient for the electrolyte $\alpha_e$ is defined as
\begin{equation}
    \ln(\alpha_e) = \frac{\mu_e - \mu_e^\theta}{R\,T}
\end{equation}
where $\mu_e^\theta$ denotes an arbitrarily chosen zero for the component $e$ and is called the {\it standard chemical potential} of $e$. The $\chi$-parameter for the solvent and the electrolyte can be obtained from $\alpha_e$:
\begin{equation}
    \chi = \frac{\ln(\alpha_e) - \ln(1-\phi)}{\phi^2}
    \label{concdependentchi}
\end{equation}
and its explicit concentration-dependence comes about by writing $\alpha_e =(x)^x \,(y)^y\,(\alpha_e^0\,m)^z$, where $x$ and $y$ are the stoichiometric coefficients of the cation and the anion, and $z = x + y$. $\alpha_e^0$ denotes the {\it mean} activity coefficient of the electrolyte, and $m$ its molality. Equation (\ref{concdependentchi}) allows one to obtain the Flory-Huggins concentration-dependent parameter if the activity coefficient is known. The scaling of $\chi$ with the quantity of ions present has been studied in~\cite{mayoral2}. The behaviour of this quantity as a function of the concentration $\zeta$ follows a power law $\chi \sim \zeta^\tau$ with characteristic scaling exponent $\tau$ dependent on the kind of salt.

Comparing Eqs. (\ref{virialfreeenergy}) and (\ref{DeltaF}), Groot and Warren~\cite{groot} proposed that the repulsive parameters $a_{AB}$ in the DPD simulation can be obtained using the $\chi$-Flory-Huggins parameter as
\begin{equation}
    \chi_{AB} = \frac{\alpha\,(2\,a_{AB} - a_{AA} - a_{BB})\,\rho}{k_B\,T}
    \label{gw}
\end{equation}
and using (\ref{gw}) and (\ref{concdependentchi}) the repulsive DPD parameter $a_{ij}$ dependent on the concentration may be obtained
\begin{equation}
    a_{ij} = a_{ii} + 3.27\,\chi_{ij}
    \label{adependentonchi}
\end{equation}
with, as before,
\begin{equation}
    a_{ii} = \frac{k_B\,T\,(\kappa^{-1} N_m - 1)}{2\,\alpha\,\rho_{DPD}}
\end{equation}
Thus, for $3$ water molecules per particle ($N_m = 3$) and a compressibility of $\kappa^{-1} \approx 16$ for water at $300^{\circ} K$ and $1$ atm, we have $a_{ww} = 78.3$.

\subsection{Temperature dependence of the DPD interaction parameters}

When the heat of mixing is given by the Hildebrand-Scatchard regular solution theory~\cite{hildebrand, scatchard, hildebrand2, barton} the $\chi_{ij}$-parameter can be obtained using the solubility parameters $\delta_i(T),\ \delta_j(T)$ for the pure components in the mixture, which themselves are temperature-dependent. We have
\begin{equation}
    \chi_{ij}(T) = \frac{v_{ij}}{R\,T}\,\left[\delta_i(T)-\delta_j(T) \right]^2
    \label{chidependentondelta}
\end{equation}
with $v_{ij}$ the partial molar volume. While this approximation is valid for non-polar components, it has been used in polar systems with reasonable success~\cite{blanks, AGGEMVT}. From Eqs.(\ref{adependentonchi}) and (\ref{chidependentondelta}) we have
\begin{equation}
    a_{ij}(T) = a_{ii}(T) + 3.27\,\frac{v_{ij}}{R\,T}\,\left[\delta_i(T)-\delta_j(T) \right]^2
    \label{adependentonT}
\end{equation}

The determination of solubility parameters is a difficult and laborious undertaking, but correlations with other physical properties of the substance in question help. For example, writing
\begin{equation}
    \delta^2 = \delta_d^2 + \delta_p^2 + \delta_h^2
    \label{solparam}
\end{equation}
where $\delta_d^2$ denotes the {\it dispersion} component of the total solubility parameter,  $\delta_p^2$ its {\it polar} component, and $\delta_h^2$ its contribution from {\it hydrogen bonding}, the dispersion component $\delta_d$ may be very well approximated by using the total solubility parameter of a {\it homomorph} molecule, i.e., a non-polar molecule most closely resembling the molecule in question in size and structure ($n$-butane is the homomorph of $n$-butyl alcohol, for example). This is because the solubility parameter of the homomorph is due entirely to dispersion forces. One still needs to determine either $\delta_p$ or $\delta_h$ (the other one is obtained by substraction from the total solubility parameter $\delta$ using Eq.(\ref{solparam}), when known), and this is done through trial and error experimentation on numerous solvents and polymers and comparing similar and dissimilar structures according to functional groups and molecular weight.

The total solubility parameter may be calculated from the cohesive energy $E_{coh}$ or, equivalently, from the enthalpy of vapourisation $H^{vap}$
\begin{equation}
    \delta_A = \sqrt{\frac{\Delta E_{coh}}{V_A^0}} = \sqrt{\frac{\Delta H^{vap}-RT}{V_A^0}}
\end{equation}
by using atomistic dynamic simulations. To do this, periodic cells of amorphous fluid structures may be constructed using regular available software such as the {\it Amorphous Cell} program of {\it Materials Studio}. The dimension of the box is specified (e.g. $25$ {\AA} on each side). Interatomic force-field interactions are set as initial conditions, and the system is evolved according to Eq.(\ref{Newton}).

The solubility parameter of a mixture of liquids is determined by calculating the volume-wise contributions of the solubility parameters of the individual components of the mixture, i.e., the parameter for each liquid is multiplied by the fraction that the liquid occupies in the blend, and the results for each component added together. In these multicomponent systems the $\chi$-parameters are calculated by pairs. If, for instance, we have a $3$-component mixture of water $w$ (or other solvent), electrolyte $e$, and an organic compound $o$, we have
\begin{eqnarray}
    \chi_{ew} &=& \frac{v_{ew}}{R\,T}\,\left[\delta_e(T)-\delta_w(T) \right]^2 \\
    \chi_{wo} &=& \frac{v_{wo}}{R\,T}\,\left[\delta_w(T)-\delta_o(T) \right]^2 \\
    \chi_{eo} &=& \frac{v_{eo}}{R\,T}\,\left[\delta_e(T)-\delta_o(T) \right]^2
\end{eqnarray}
and, in fact, taking the square root of of any two of these equations (the first two, say), adding them together, and assuming $v_{ew}=v_{wo}=v_{eo} \equiv v_{m}$, we can have a very good estimate for the third
\begin{equation}
\left[ \sqrt{\chi_{ew}} + \sqrt{\chi_{ew}} \right]^2 = \frac{v_{m}}{R\,T}\,\left[\delta_e(T)-\delta_o(T) \right]^2 \equiv \chi_{eo}
\end{equation}

Although we have assumed heretofore that DPD particles mix randomly, and that the particles of a given type are indistinguishable, this model predicts very well the major trends in the behaviour of real polymer solutions and is used to predict new behaviour in polymers in current research~\cite{langmuir, mayoralSpringer2, mayoral1, mayoral2}.

\section{Applications}

\subsection{Interfacial tension}

Interfacial tension arises from the contact between immiscible fluids. It is a measurement of the cohesive (excess) energy present, arising from the imbalance of forces between molecules at the interface. This excess energy is called {\it surface free energy} and is a measurement of the energy required to increase the surface area of the interface by one unit. Equivalently, it may be quantified as a force/length measurement: the force which tends to minimise the surface area. Interfacial tension plays an important role in the formation of colloids or emulsions: as each phase tries to maintain as small an interface as possible, they do not easily mix. Similarly, it is important for the dispersion of insoluble particles in a liquid medium, the penetration of molecules through membranes, adsorption, and stability.

The measure or otherwise determination of the interfacial tension then allows us to study the hydrodynamics and morphology of multiphase systems, and this in turn is a most important aspect of the understanding of natural processes and of product design.

The conservative force allows us to calculate the average kinetic energy $E_k$ via the virial theorem
\begin{equation}
    2\,\langle E_k \rangle = -\sum_{i=1}^N\,\langle F_i{}^C \cdot r_i\rangle
\end{equation}
where $F_i{}^C$ is the total conservative force on DPD particle $i$: $F_i{}^C = \sum_{j=1}^N F_{ji}{}^C$ with $F_{ji}{}^C$ the force applied by particle $j$ on particle $i$; and from $\langle E_k \rangle$ we may calculate the fluid pressure tensor
\begin{equation}
    P_{\alpha\beta} = \frac{1}{V}\,\left( \sum_{i=1}^N m_i\,v_{i\beta}\,v_{i\alpha} + \sum_{i=1}^N F_{i\beta}\,\alpha_i \right)
    \label{pressuretensor}
\end{equation}
Here, $m_i$ is the mass of particle $i$ (which we set equal to $1$ in DPD-units) and $v_{i\alpha}$ is the $\alpha$-component of the velocity of particle $i$ in the volume $V$; similarly, $F_{i\beta}$ is the $\beta$-component of the force $F_i$ on particle $i$, $\alpha_i$ is the $\alpha$-coordinate of particle $i$, etc. Eq.(\ref{interfacial}) may then be used to calculate directly the interfacial tension $\gamma$ at the volume boundary, with $\gamma = (k_B\,T/r_c)\,\gamma^\ast$.

$\gamma$ is dependent on temperature. From the mechanical work needed to increase a surface area, $dW = \gamma\,dA$, we have
\begin{equation}
    \gamma = \left( \frac{\partial G}{\partial A} \right)_{T,P,n}
\end{equation}
with $G$ the Gibbs free-energy and $A$ the surface area. As all spontaneous thermodynamic processes follow $\Delta G < 0$, it is easy to understand why the liquid tries to minimise its surface area. From its definition, $G = H - T\,S$ with $H$ the enthalpy and $S$ the entropy of the system. Thus
\begin{equation}
    \left( \frac{\partial \gamma}{\partial T} \right)_{A,P} = -\frac{S}{A}
\end{equation}
so that the normal behaviour of $\gamma$ is to decrease with temperature.

\begin{figure}
\begin{center}
\scalebox{0.7}{\includegraphics{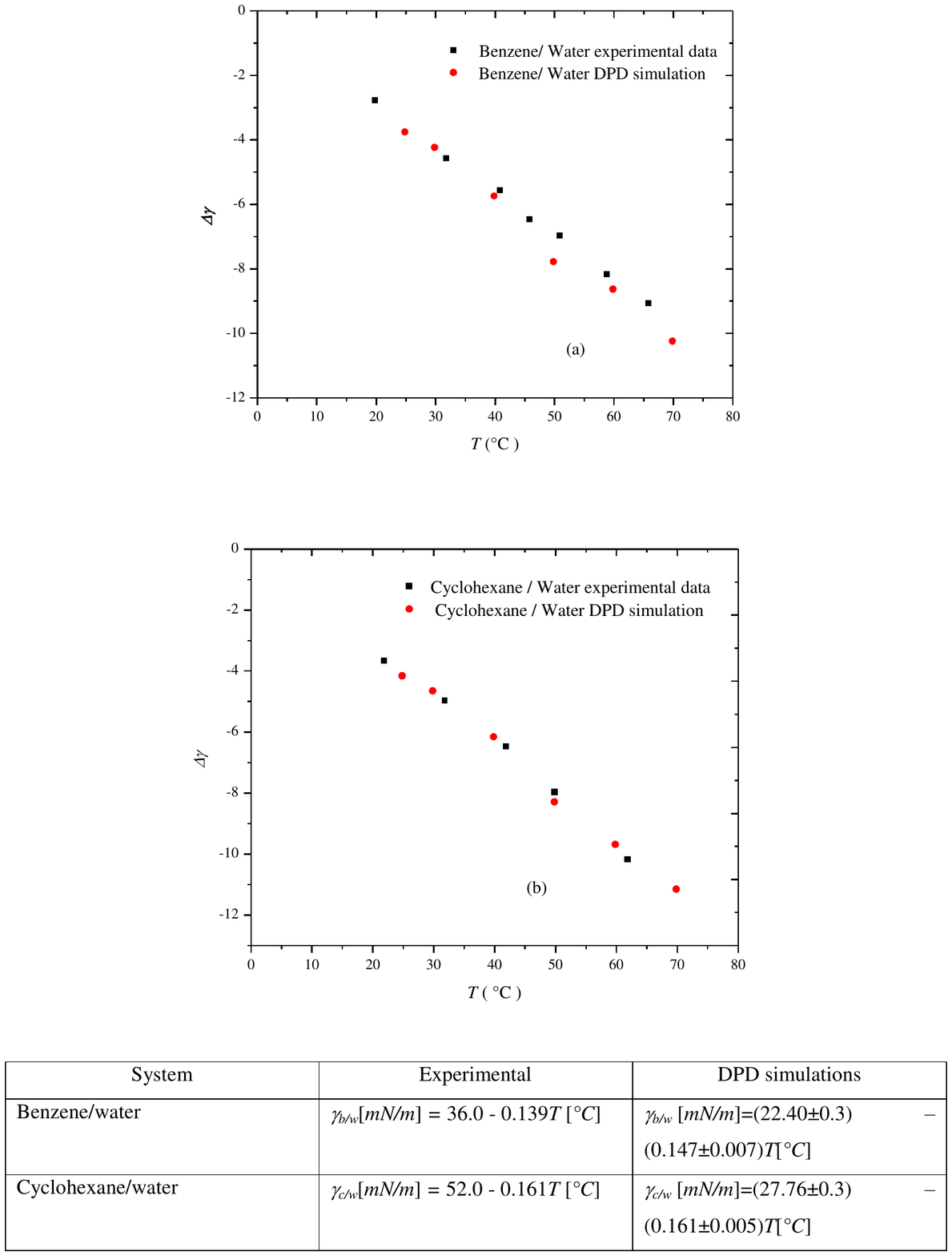}} \\
\end{center}\caption{Interfacial tension for Benzene/Water and Ciclohexane/Water mixtures at different temperatures using DPD simualtions at different temperatures.} \label{ITT}
\end{figure}

Results concerning the study of the interfacial tension between
immiscible mixtures such as benzene/water and ciclohexane/water at
different temperatures, using the parametrisation mentioned above and
performing DPD simulations, can be found in~\cite{AGGEMVT}. These reproduce
the experimental data as shown in Figure~\ref{ITT}, and confirms that the parametrisation via the use of solubility parameters at different temperatures to obtain the
repulsive DPD parameters $a_{ij}$ as functions of $T$ is appropriate for introducing the effect of temperature in DPD simulations.

\begin{figure}
\begin{center}
\scalebox{0.40}{\includegraphics{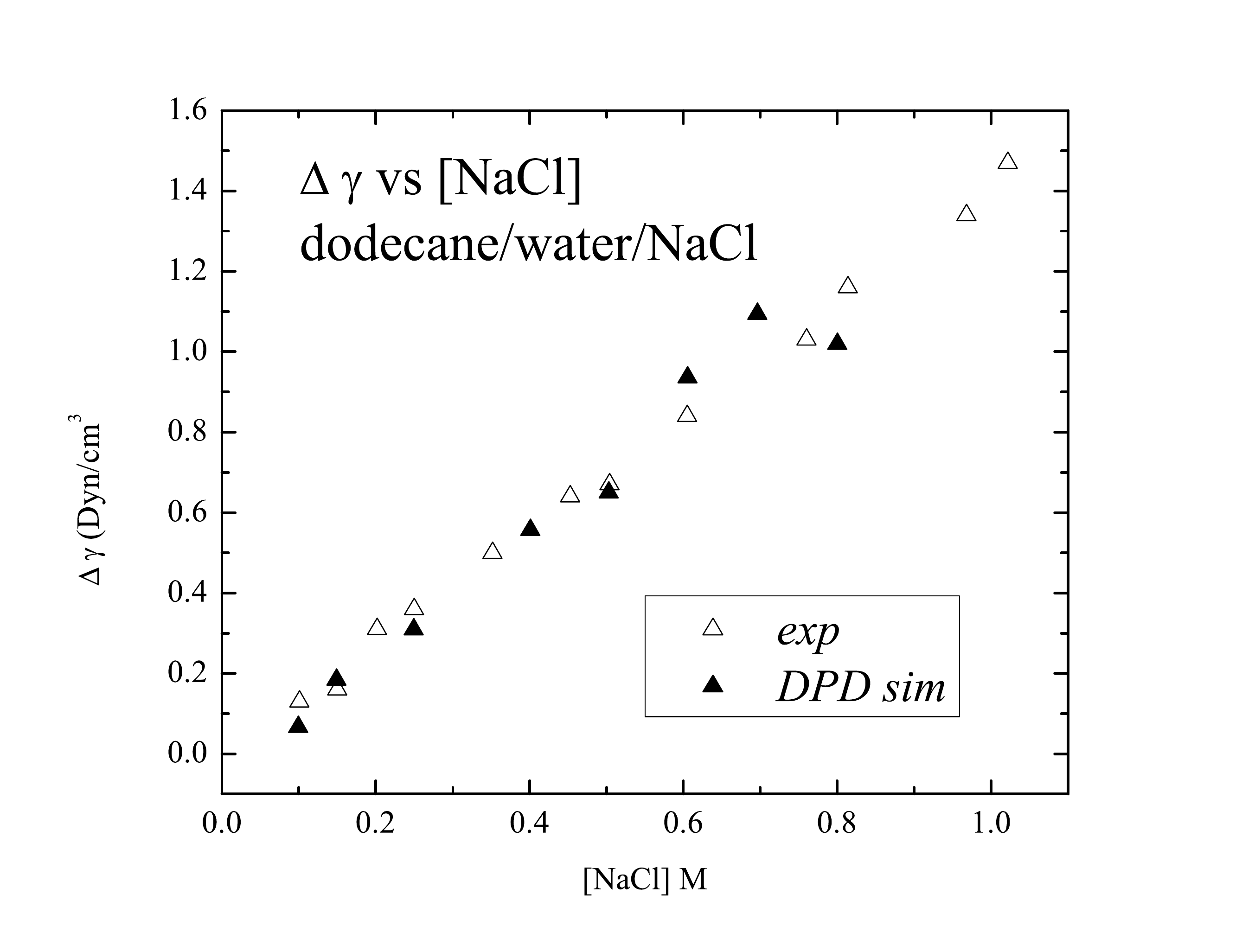}} \\
\scalebox{0.40}{\includegraphics{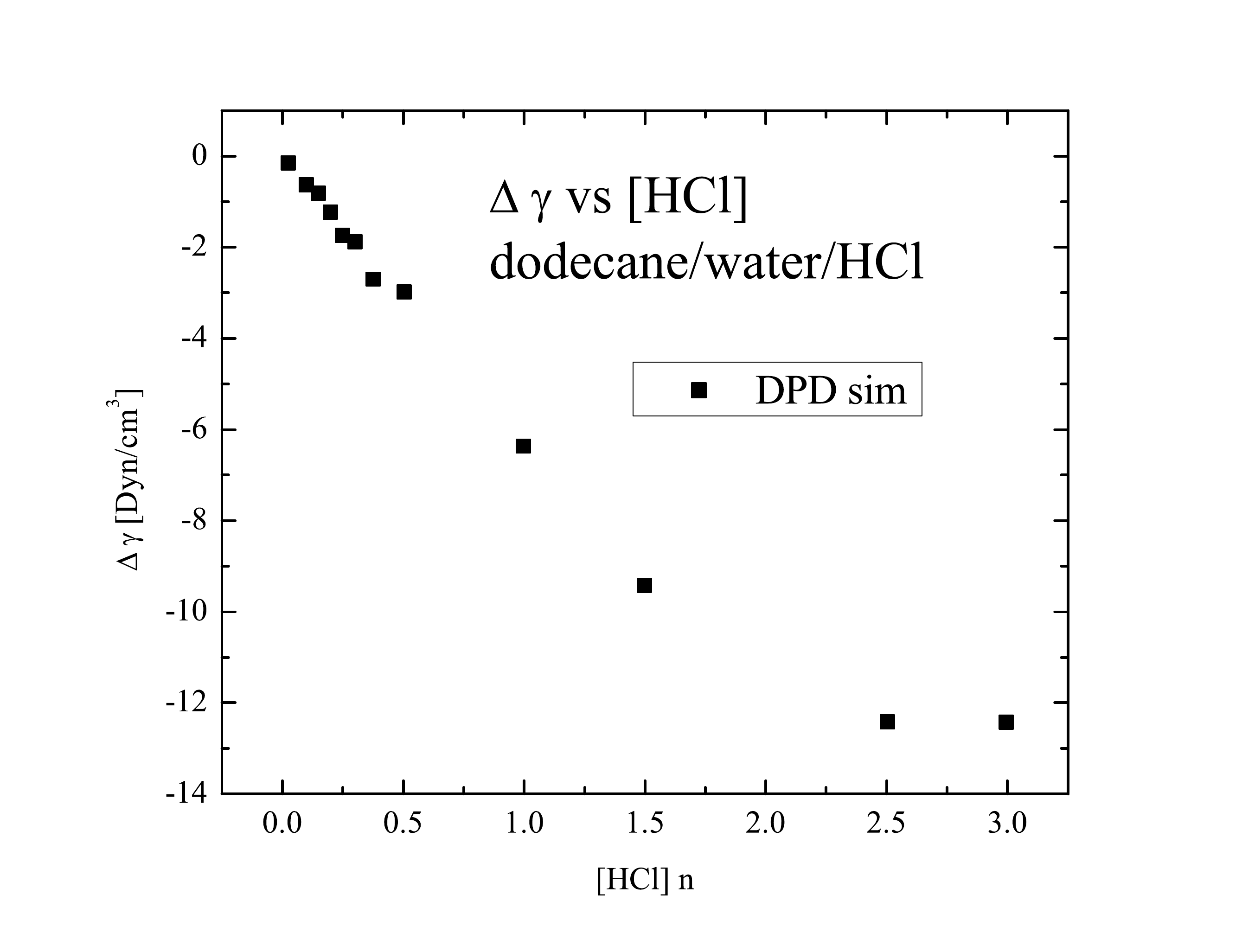}}
\end{center}
\caption{Interfacial tension experimental data are showed with withe triangles and DPD simulations results are showed in black triangles for $n$-dodecane--water with $n[NaCl] $
(top) and $HCl$ (bottom) added.}
\label{IT}
\end{figure}

Additionally, the interfacial tension between two species will
change when an electrolyte is added at different concentrations,
since the cohesive forces between neighbouring molecules will be
altered. Its behaviour with concentration will depend strongly on the
type of electrolyte. Figure~\ref{IT} (top) shows the behaviour of the
interfacial tension $\gamma^\ast$ between $n$-dodecane and water
with sodium chloride $NaCl$ added, obtained by DPD electrostatic
simulations. In this figure $[NaCl]\ M$ denotes the number of DPD ions
added as molar concentration. The increase with salt concentration is expected, and the
same behaviour is observed when several other inorganic salts are
added~\cite{mayoral2}. The opposite behaviour is observed, however,
when hydrochloric acid ($HCl$) is added to the same solvent mixture,
as shown in Fig.~\ref{IT} (bottom). This shows that not only the ionic charge is important but also the kind of ionic species in the mixture, because it modifies the
chemical potential.

\subsection{Adsorption isotherms}

The adsorption of polymers onto different surfaces has been the subject of many theoretical and experimental studies. Specifically, the adsorption of polyelectrolytes is a topic of extensive concern because of its practical applications. Many surfactants and additives are polyelectrolytes, and they must be adsorbed with great selectivity on different surfaces in order to have a good performance. This phenomenon is observed in different fields such as water purification where the adsorption of polyelectrolytes could produce flocculation. Other critical examples are emulsifiers in the food and pharmaceutical industries, as well as complex polyelectrolytes for medical science applications, among others. In order to have a good understanding of this phenomenon, more precise information about the conformation of polyelectrolytes adsorbed on a surface and living in the surrounding medium is important. Few theoretical studies have been developed to describe polyelectrolyte adsorption and experimental studies are laborious. For this reason numerical simulation seems a very good alternative. DPD simulations can reproduce the behaviour of this kind of systems but some considerations must be taken.

By construction, the DPD dynamics keep the number of particles $N$, the cavity volume $V$, and the temperature $T$ constant. For adsorption isotherms one needs the chemical potential
\begin{equation}
    \mu_i = \left( \frac{\partial U}{\partial N_i} \right)_{S,V,N_{j\neq i}}
\end{equation}
fixed; i.e., one needs to work in a Grand Canonical Ensemble $(\mu,\,V,\,T)$. This may be achieved by using a hybrid DPD--Metropolis Criterion (DPD/MC). In this, after the usual DPD dynamics, where the initial $\mu(t_0)$ drifts to $\mu(t)$ one performs a certain number of cycles of particle exchange with the virtual bulk that will return the chemical potential to its initial value $\mu(t_0)$, and calculates the final energy of the system: if equal or lower than the initial energy, the exchange cycle is accepted; if higher it is rejected and a new exchange cycle is performed. This is followed by another iteration of DPD dynamics together with particle exchange cycle, and so on. By generating separate simulations for different polymer concentrations in this manner, one may calculate the density profile $\rho(z)$ in a box of length $L_z$, and from it the adsorption $\Gamma$ as

\begin{equation}
    \Gamma = \int_0^{L_z}\,\left[\rho(z) - \rho_{\text{bulk}}\right]\,dz
\end{equation}

\begin{figure}
\begin{center}
\scalebox{0.6}{\includegraphics{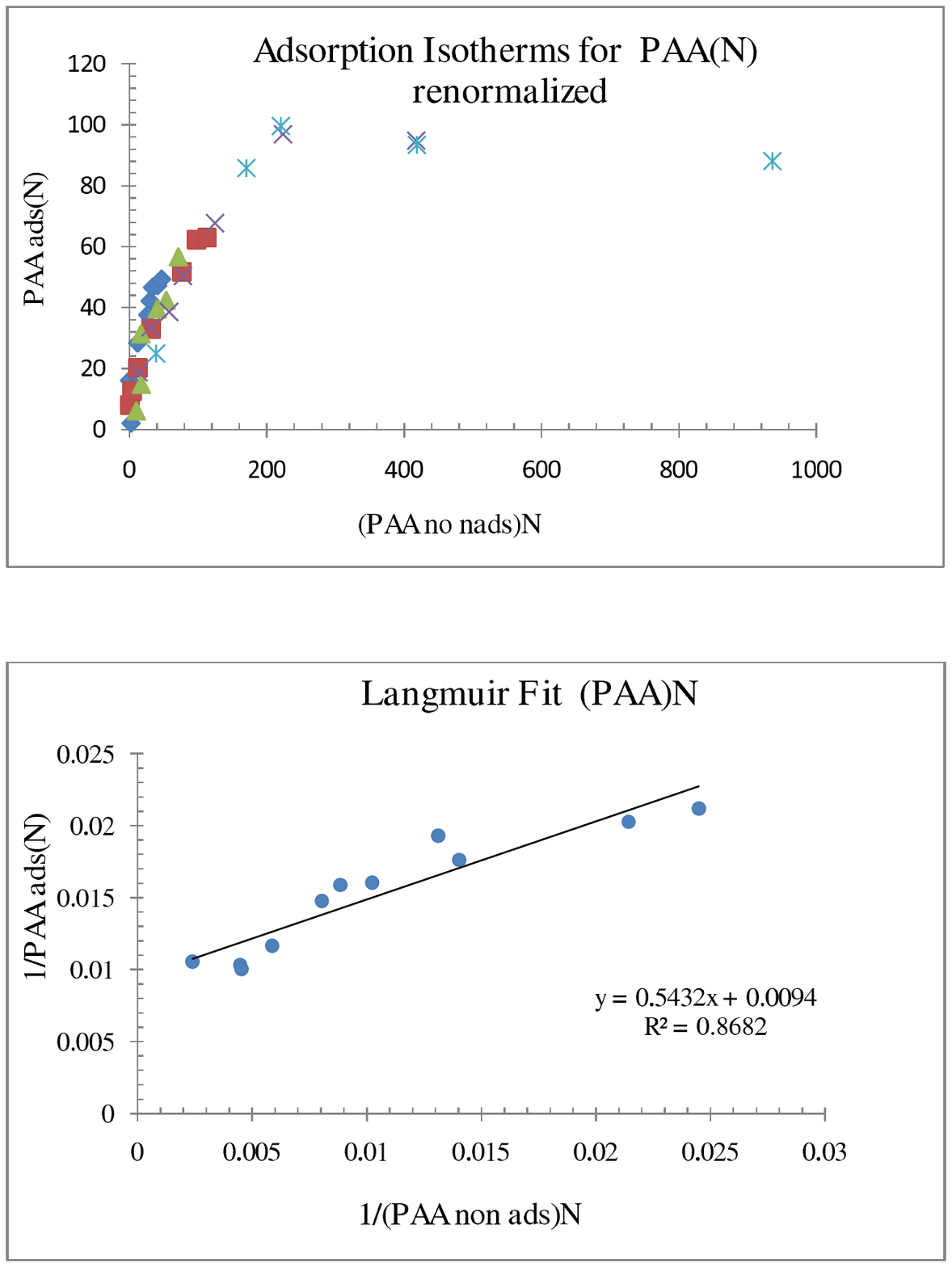}}
\end{center}
\caption{Adsorption isotherm for $PAA$ on $TiO_2$ via electrostatic
DPD simulation.} \label{IAPAA}
\end{figure}

Adsorption isotherms have been calculated performing DPD simulations in this
manner~\cite{mayoral3} and checked to coincide with
experimental determinations~\cite{hulden, esumi}. As an example, Figure~\ref{IAPAA} presents the results for the simulation of the adsorption of polyacrylic acid (PAA) on $TiO_2$ surfaces. PAA was mapped considering each DPD bead as one monomeric unit ($-CH_2-COOH$). The repulsive $a_{ij}$ parameters were obtained according Section~\ref{parametrisation}. The number of independent adsorbed vs non-adsorbed DPD beads is presented. If we assume that only one layer is adsorbed on the surface and all adsorption positions are equivalent, we can extract the maximum concentration at equilibrium and the adsorption-desorption constant, which is given by the Langmuir isotherm. We consider that the ability of one monomeric unit to be adsorbed onto one site of the surface is independent of occupied sites next to it. The expression for this kind of adsorption model is given by the Langmuir isotherm expressed by

\begin{equation}
   \frac{1}{\Gamma} = (\frac{1}{\Gamma_M} + \frac{1}{\Gamma_M KC})
\end{equation}
where $K = K_a/K_d$ and $C$ is the concentration in the bulk, $\Gamma$ is the adsorbed quantity and $\Gamma_M$ is the maximum adsorbed quantity. A linear fit for this isotherm is shown in Figure~\ref{IAPAA} and it could be seen that $1/\Gamma_M = 0.8829$, $\Gamma_M = 1.13257$ and $K = 6.4476$. Taking into account a surface area for $TiO_2$ of $30.22 m^2/g$, results in $\Gamma_M = 7.987 (mg PAA/mg TiO_2)$. This value corresponds well with the experimental data reported in~\cite{hulden,mayoral3} of $\Gamma_M = 6.96 (mg PAA/mg TiO_2)$.

\subsection{Disjoining pressure}

\begin{figure}
\begin{center}
\scalebox{0.5}{\includegraphics{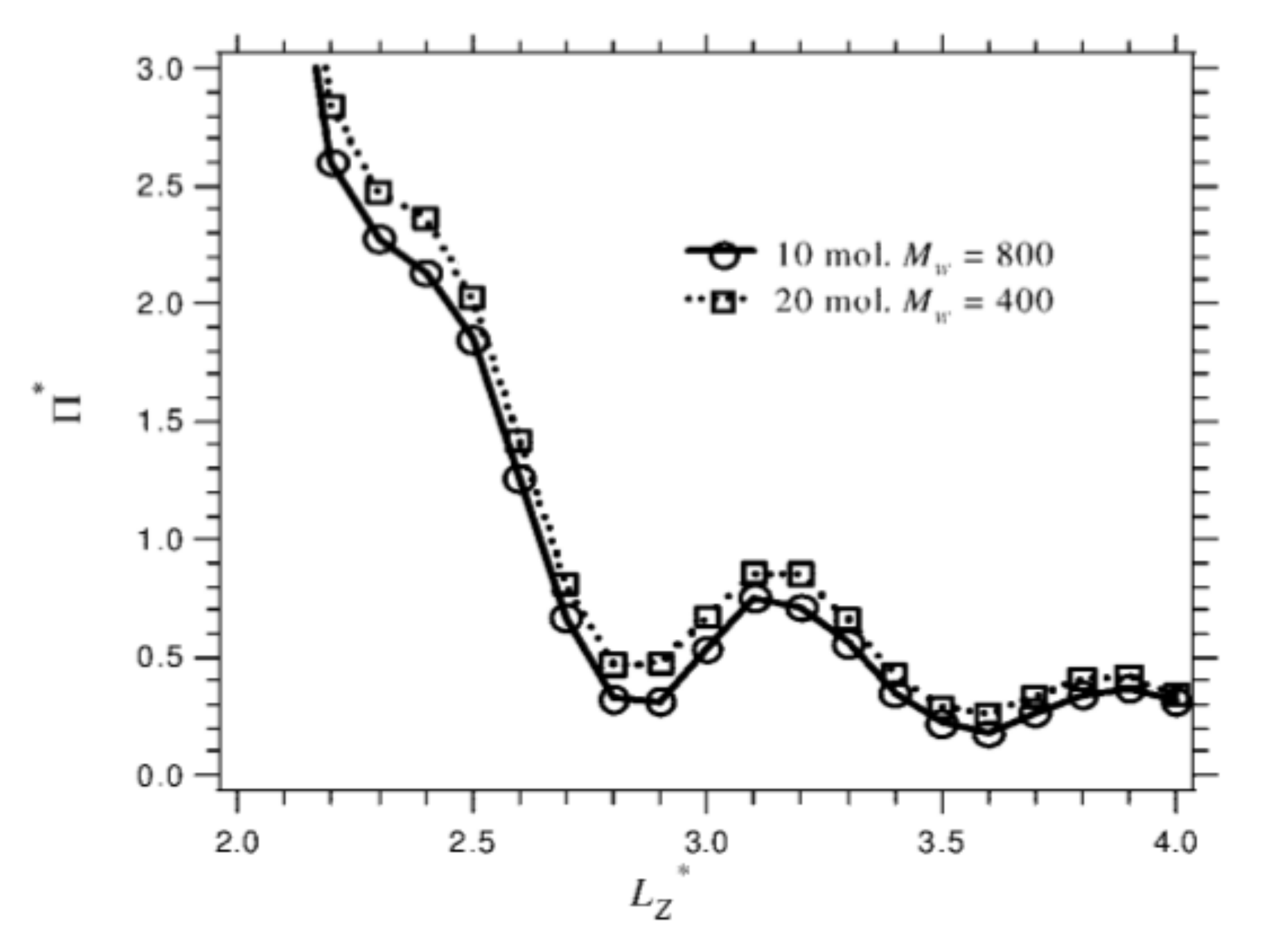}}
\end{center}
\caption{Disjoining pressure isotherms for $20$ molecules of short-chain $M_w=400$, $vs.$ $10$ molecules of long-chain $M_w=800$ surfactant molecules of the PEG-type.}
\label{DP}
\end{figure}

Colloid stability strongly depends on the {\it disjoining pressure}. For a confined fluid, the pressure component perpendicular to the confining walls $P_N$ is different from the unconfined bulk pressure $P_{bulk}$. This differential pressure relative to the bulk, which is a function of the separation $L_z$ between the parallel walls is called ``disjoining pressure''. For a wall perpendicular to the $z$-direction
\begin{equation}
    \Pi(L_z) = P_{zz}(L_z) - P_{bulk}
\end{equation}
While $P_{bulk}$ is obtained from the average of the diagonal components of the pressure tensor (cf. Eq.(\ref{pressuretensor}) above), the pressure normal to the wall is calculated from the $zz$-component, averaged over the length $L_z$ of the simulation box in the direction perpendicular to the walls. Equivalent expressions are used for $P_{xx}$ and $P_{yy}$. The disjoining pressure is a measure of the force, per unit area, needed to bring $2$ particles (or a particle and a substrate) together, thus providing a criterion for stability. It has been calculated~\cite{langmuir} for different types of surfactants (those that graft at one end onto a substrate, and those that can adsorb onto the substrate along their full length thus acting as surface modifiers) and for different substrates. The results show that the greater stability attained is not a consequence of greater molecular weight of the dispersant species itself, as so often misinterpreted, but rather of greater molecule mobility. I.e., the entropic gain of having monomers with more mobility to sample the configurational space than polymers (at the same monomer concentration) is the leading mechanism responsible for the higher values of disjoining pressure. This is shown in Figure~\ref{DP} for a surface-modifying polymer. In this figure we observe the typical oscillations in $\Pi$ present in confined fluids~\cite{israelachvili}. While maxima in $\Pi$ correspond to more stable thermodynamic configurations, minima represent regions of instability. In this case molecules with a molecular weight $M_w = 400$ were considered, corresponding to $7$ DPD-particles joined by springs. Having $20$ such molecules present amounts to having 140 monomeric units, a concentration that can also be achieved by considering $10$ polymeric molecules of $M_w=800$ of the same chemical type.

Polyethylene glycol (PEG) of $M_w=400$ and $M_w=800$ were used for the results in Figure~\ref{DP}, with a DPD-particle volume of $90$ \AA$^3$ which can accommodate $3$ water molecules. The repulsive wall interaction parameter was chosen as $a_{w-monomer}= 60$ when the particle interacting with the wall was a monomer of the polymer molecule, and as $a_{w-sol}= 120$ for solvent molecules. For particles of the same species we took $a_{ii}=78.0$ and for particles of different species $a_{ij}=79.3$. Our choices reproduce isothermal compressibility of water at room temperature, and promotes polymer adsorption onto the substrate over solvent adsorption. For the spring constant in the polymer DPD-particles we took $k=100$ with an equilibrium distance of $r_{eq}=0.7$. The temperature was kept constant at $T=300^\circ K$.

We may observe that shorter polymers are better as dispersants when compared with longer ones at the same monomer concentration. If we multiply the dimensionless $\Pi^\ast$ depicted in the figure by $k_BT/r_c{}^3$ (cf. Eq.(\ref{virialpressure})), the disjoining pressure for short polymers can be up to $4.5\, \times\, 10^5$ Pa larger than that for the longer chains at certain wall separations. Stability via surface modification is then much better attained through the use of monomeric species, than through polymer chains. The same behaviour is found for grafted polymers (cf.~\cite{langmuir} for details).

\subsection{Radius of gyration}

The radius of gyration is a measure of the size of an object of arbitrary shape. For a polymer chain in solution, however, this is not a very useful definition as it can take many different configurations. One may calculate a {\it root mean square end-to-end distance} $R_{RMS}$ of the chain as
\begin{equation}
    R_{RMS}{}^2 = \langle\,(r_N - r_0)^2\,\rangle
\end{equation}
where we have denoted by $r_i,\ (i=0,\,1,\,...,\,N)$ the positions of the chain joints (i.e., the two ends of the $i$-th bond are $r_{i-1}$ and $r_i$). A more useful quantity, however, is the {\it radius of gyration} $R_g$ of the chain, given by
\begin{equation}
    R_g{}^2 = \langle \frac{1}{N+1}\,\sum_{i=0}^N (r_i - r_{CM})^2 \rangle
\end{equation}
where $r_{CM} = \frac{1}{N+1}\,\sum_{i=0}^N r_i$ is the centre of mass of the chain. Loosely speaking, the chain occupies the space of a sphere of radius $R_g$, i.e., it intuitively gives a sense of the size of the polymer coil. Note that $mR_g{}^2$ (with $m$ the mass of the polymer molecule) is the moment of inertia of the molecule about its centre of mass, and that we can write the equation above as
\begin{equation}
    R_g{}^2 = \frac{1}{2} \langle \frac{1}{(N+1)^2}\,\sum_{i,j=0}^N (r_i - r_j)^2 \rangle
\end{equation}
which is useful since it allows us to calculate the radius of gyration of the molecule by using the mean square distance between monomers, without calculating $r_{CM}$. Note also that we have used averaging in all the equations above; this is because the possible chain conformations are numerous and constantly change in time, thus we understand the radius of gyration as a mean over time of all the polymer molecules, which by ergodicity principles we calculate as an ensemble average.

The radius of gyration can be easily determined experimentally through light scattering or other alternative methods (neutron scattering, etc.), allowing one to check a theoretical model against reality, and this is what makes it an interesting quantity of study. It has been extensively studied for neutral polymeric species but, as the presence of charges completely changes the possible configurations of the molecules in solution, it is interesting to study the behaviour of $R_g$ in a polyelectrolyte.

\begin{figure}
\begin{center}
\scalebox{0.40}{\includegraphics{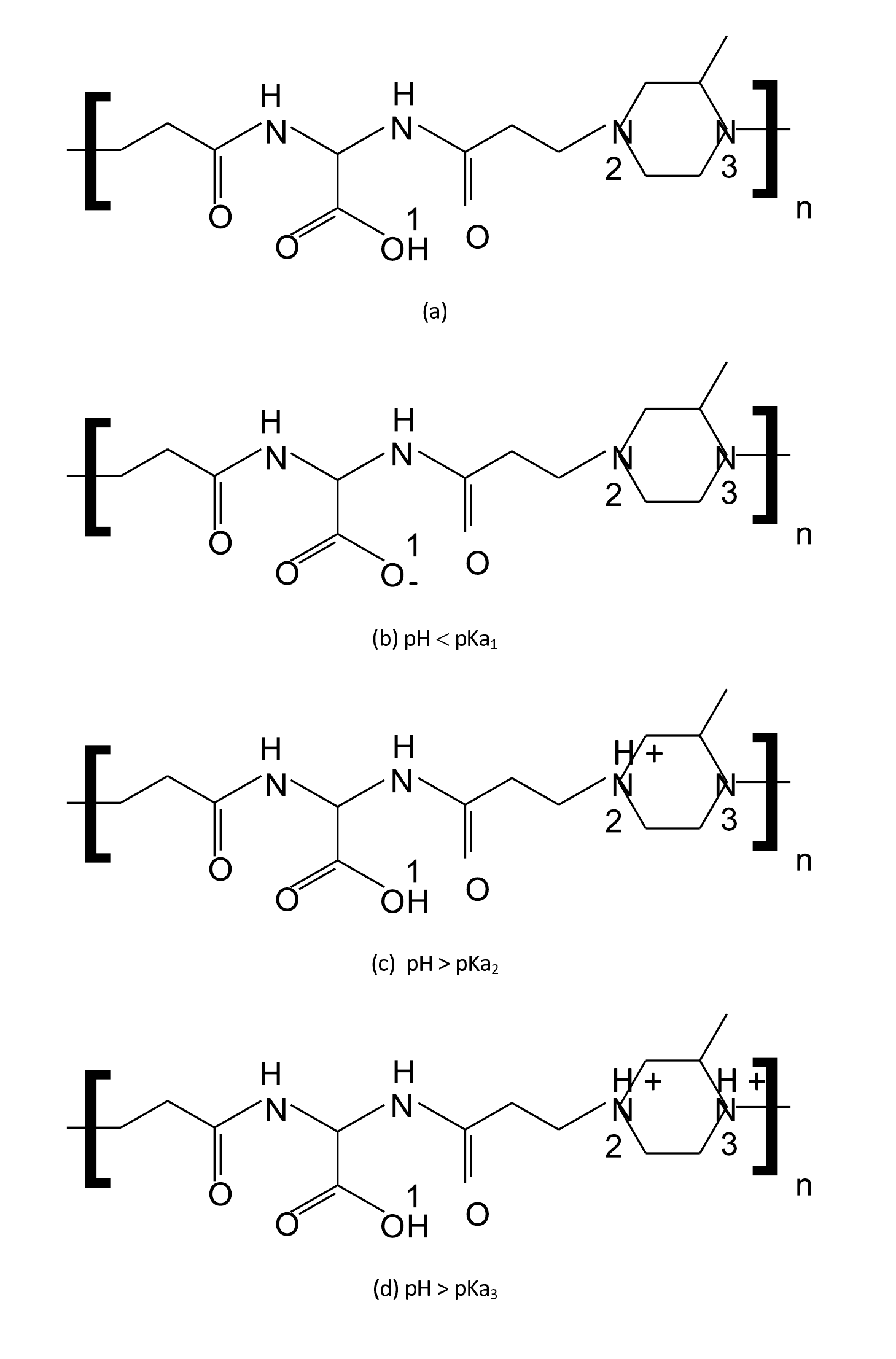}}
\caption{Neutral and ionised structures of bio-polymer PAA ISA 23.}
\label{ISA23}
\end{center}
\end{figure}

One interesting problem is the pH-dependent conformational change of
some biopolyelectrolites, because it affects directly the mechanism
of action in different situations. An example of this is the
poly(amidoamine) (PAA) which is used as endosomolytic biopolymer for
intracellular delivery of proteins and genes. Bio-responsive
behaviour of these kinds of compounds is related with the structure
and conformation in the medium, which could be estimated by the radius
of gyration. This is modified by pH and ionic strength effects.
Experimental studies of small-angle neutron scattering (SANS) have
been published in order to illustrate the pH-dependency and
conformational change of PAA ISA 23~\cite{ISA23}.
Linear poly(amidoamine) polymers (PAAs) have amido- and
tertiary amino-groups along the main polymer, which gives rise to an
interesting pH-dependent conformational change and thus offers a
perfect prospect for devising polymers that present membrane
activity at low pH. The neutral structure of this biopolymer is
shown in Figure~\ref{ISA23}(a).

The molecular weight of ISA23 is $16 500$ g/mol and it has three $pKa$'s:
$pKa_1 = 2.1$, $pKa_2= 7.5$ and $pKa_3 = 3.3$. For this reason, the
molecule could be in three different ionisation forms as illustrated in
Figures \ref{ISA23}(b), \ref{ISA23}(c), \ref{ISA23}(d).

\begin{figure}
\begin{center}
\scalebox{0.35}{\includegraphics{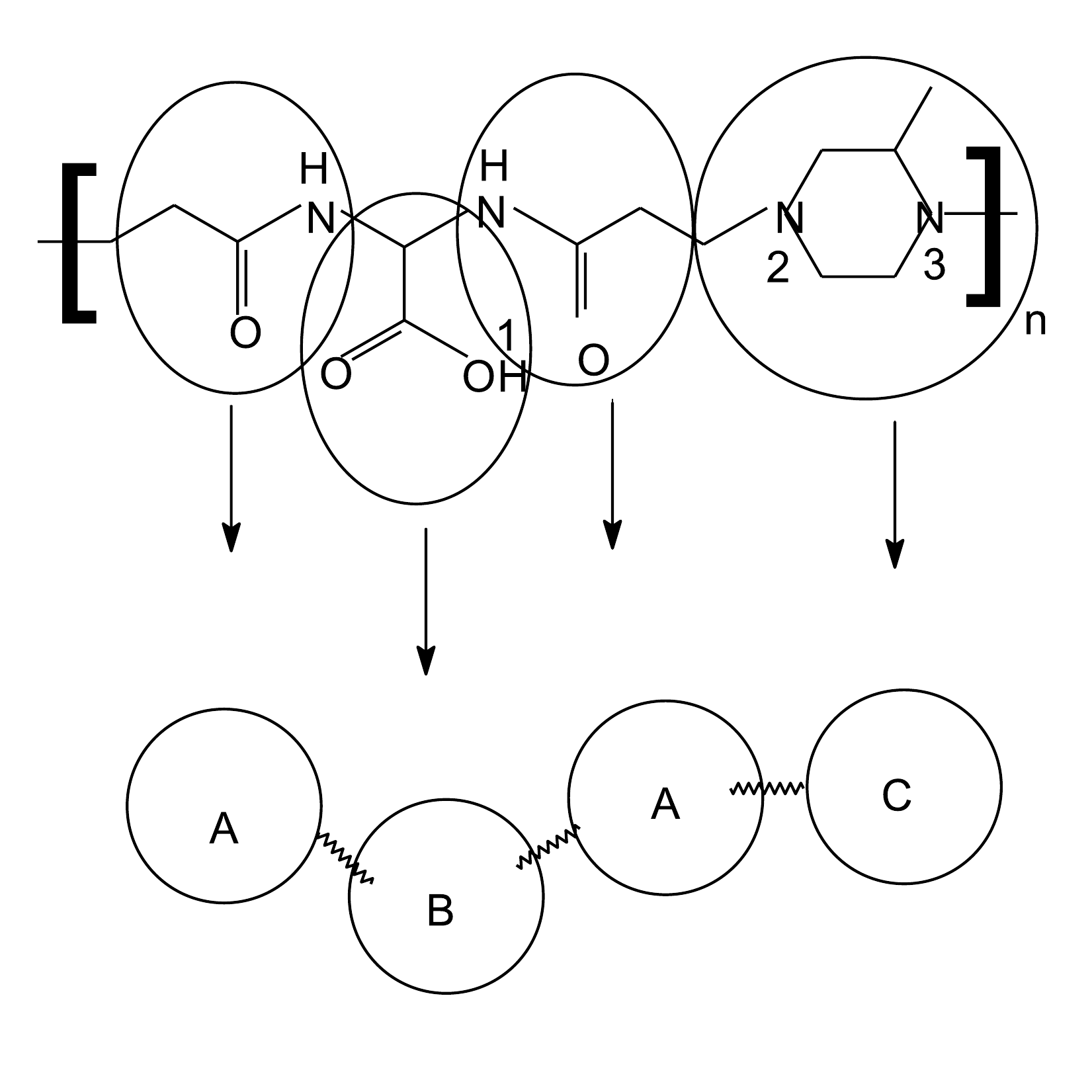}}
\caption{Mapping of PAA ISA 23 for DPD simulations.}
\label{ISAMAPEO}
\end{center}
\end{figure}

Electrostatic DPD simulations have been performed~\cite{mayoral4} in order to study the
radius of gyration of this molecule and compare with
experimental data reported. The mapping used is showed in Figure~\ref{ISAMAPEO}. It was established by taking into account the molar volume of each
segment or monomeric unit, and considering the volume of each DPD bead
as $3 V_w$ where $V_w = 30 {\AA}^3$ is the molar volume of one water molecule.

ISA23 could be considered as a weak poly-acid
and the pH could be modelled considering its ionisation degree over
the polymeric structure. Partial charges are introduced over the
molecule considering that the B-DPD bead (see Figure~\ref{ISAMAPEO}) could be
neutral or have a charge of $1^-$. The C-DPD beads could be neutral or have a positive charge of $1^+$ or $2^+$ depending on the pH of
the medium according to the acid-base equilibrium given by

\begin{equation}
pH  =
	\begin{cases}
\vspace{0.07in}
\log\left[ \frac{\theta}{1-\theta} \right] + pKa_1,
&\text{$pH < pKa_1$} \\
\vspace{0.07in}
\log\left[ \frac{\theta}{1-\theta} \right] + pKa_2,
&\text{$pKa_1 < pH < pKa_2$} \\
\log\left[ \frac{1}{1-\theta} \right] + pKa_3,
&\text{$pKa_2 < pH < pKa_3$}
	\end{cases}
\label{PH}
\end{equation}

\noindent where $\theta$ is the ratio between the number $N^-$ of protonated -
deprotonated monomeric units and the total number $N$ of monomeric
units, and $pKa_i$ is the acid-base equilibrium constant. The
variation of pH at constant ionic strength makes available the
control of the partial charge over the macromolecule. The DPD
parameters $a_{ij}$ are calculated as described in Section~\ref{parametrisation}
using the solubility parameters obtained by molecular
simulation. Ionic strength was fixed to $0.1 M$ and the pH was varied according
to equations~\ref{PH}.

\begin{figure}
\begin{center}
\scalebox{0.4}{\includegraphics{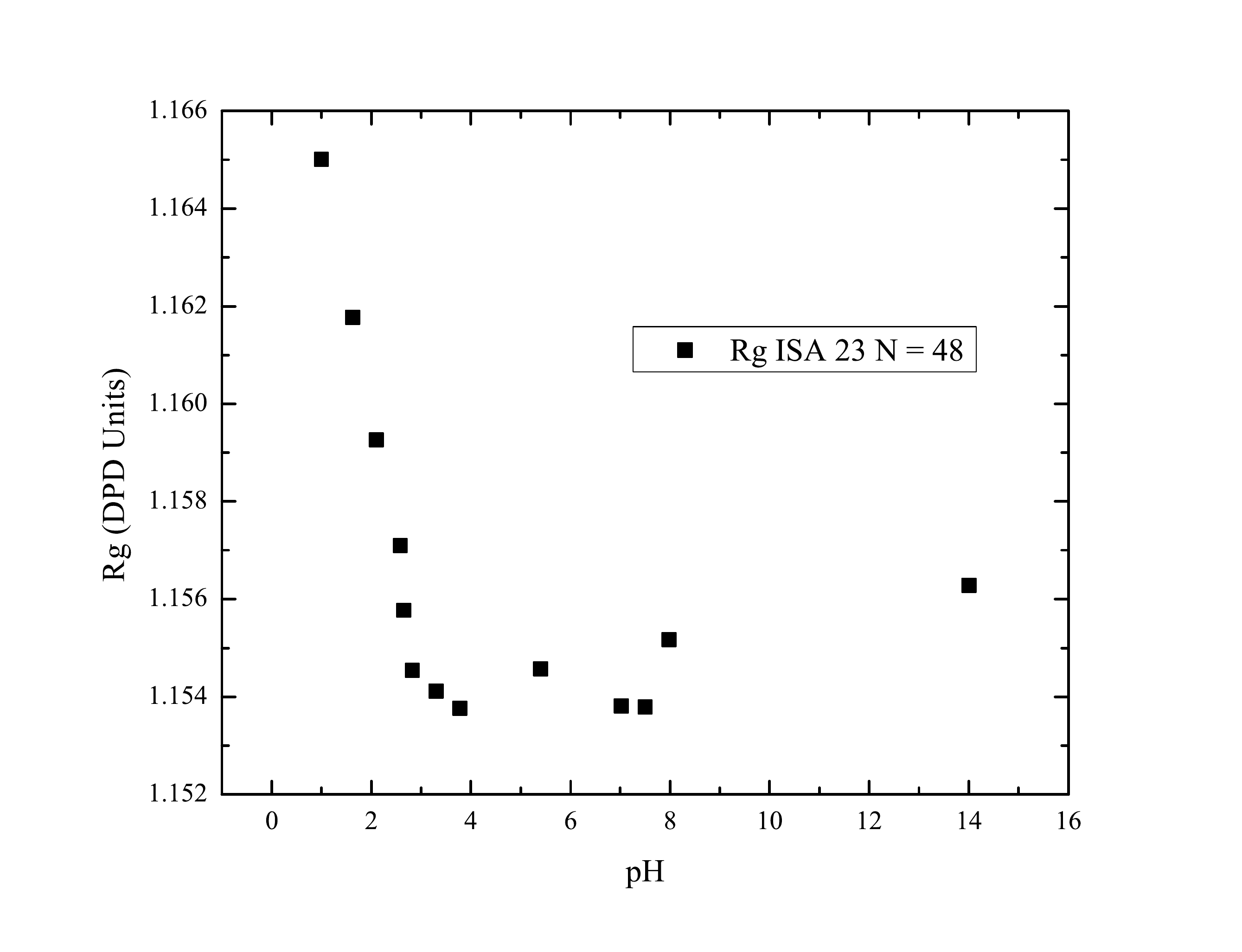}} \quad
\scalebox{0.4}{\includegraphics{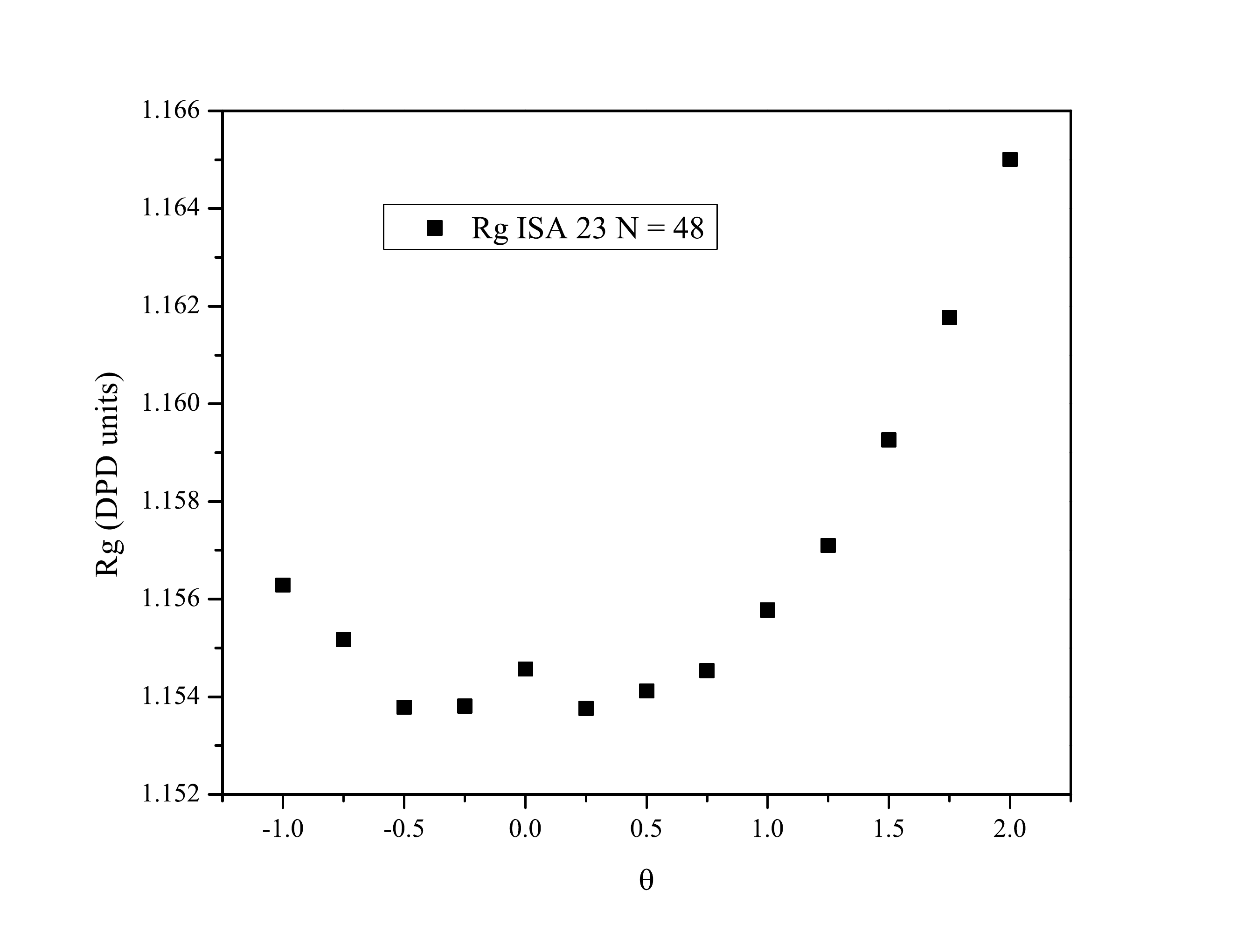}}
\caption{Top: $R_g$ vs pH for PAA ISA 23. Bottom: $R_g$ vs $\theta$ for PAA ISA 23.}
\label{rgvspHISA}
\end{center}
\end{figure}

Performing electrostatic DPD simulations at different pH's, the
mean radius of gyration was calculated for $25$ blocks of $10 000$
steps. The size of the system was $L_x = L_y = L_z = 8.5$. Also, $\gamma =
1.6$ and $\sigma = 3$. PAA ISA 23 was represented by $48$ DPD beads
joined by springs with $k = 2$. The results, as a function of pH and of $\theta$, are shown in Figures~\ref{rgvspHISA}.

\begin{figure}
\begin{center}
\scalebox{0.80}{\includegraphics{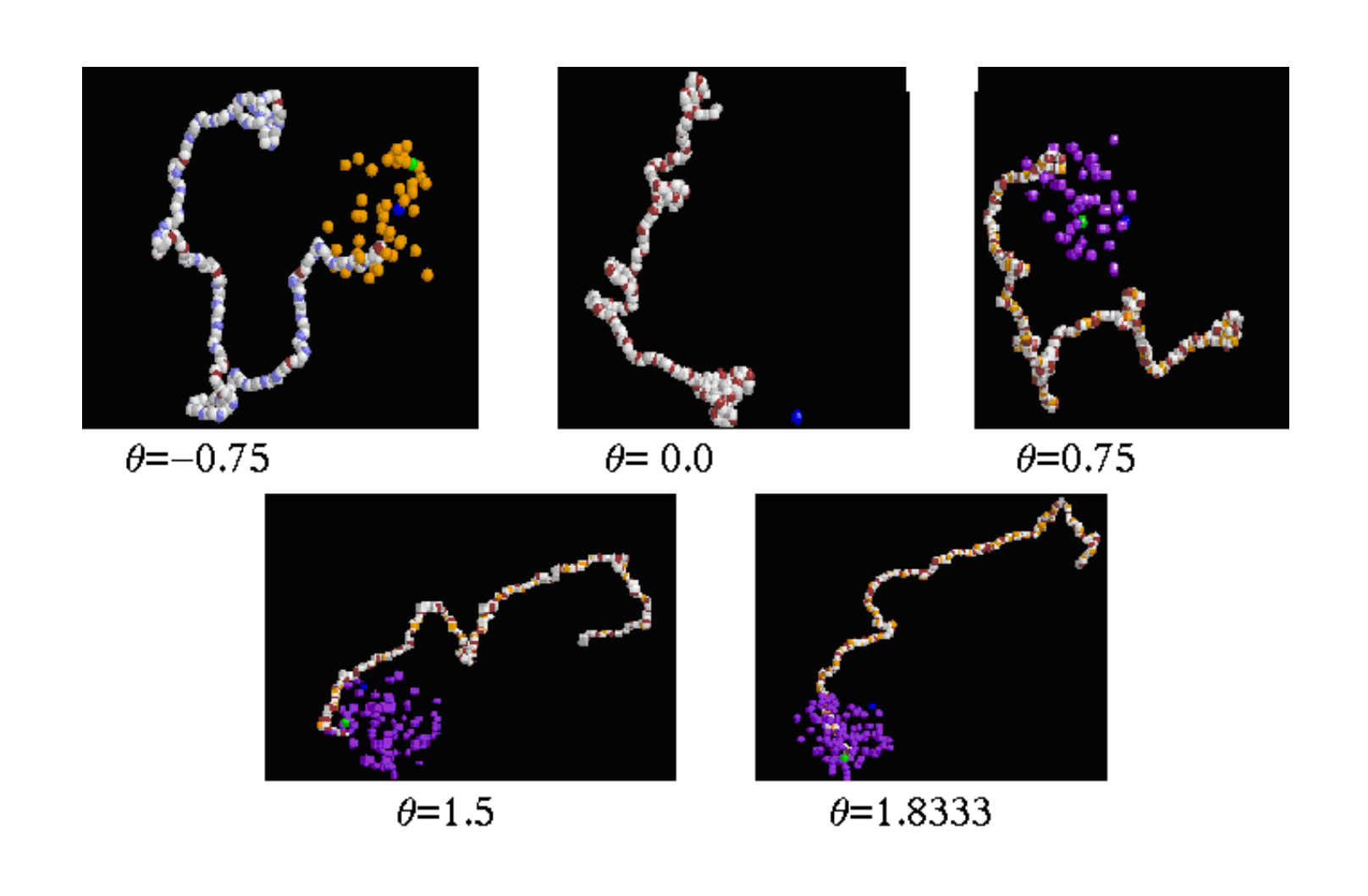}}
\caption{Conformation of PAA ISA 23 as a function of $\theta$.}
\label{simISA}
\end{center}
\end{figure}

According with these simulations, the PAA ISA 23 radius of gyration
increases to a maximum when the pH decreases. At high pH, and
therefore high ionic strengths (because of the counter-ions present
in the system), the polymer is negatively charged and adopts a
rather compact structure.  The conformation is shown in Figure~\ref{simISA}
showing how the negative counter-ions (violet beads in the figure)
are distributed near the extreme of the polymer where the amide group
is located and the internal structure is extended at low pH ($\theta = 1.8333$).
At high pH ( $\theta = - 0.75$) the positive
counter-ions (orange DPD beads in the figure) are around the carboxyl
extreme.

Experimental data reported~\cite{ISA23} shows a very similar but more complex
equilibrium in the system: with decreasing pH, the PAA ISA 23 radius of
gyration increases to a maximum around pH = 3, after which value a decreasing $R_g$
is observed when the pH is increased. At high pH the polymer is
negatively charged and presents a pretty compact structure
presumably. At low pH, the coil again collapses, and the author suggests
that this is almost certainly due to the effects of the high ionic
strength; this latter behaviour is not observed in Figure~\ref{rgvspHISA} probably because the ionic strength was fixed at $1\ M$.

\section{Scaling}

Scaling and universality are two amazing properties that
collectively have generated the modern theory of critical behaviour
appearing in different areas of modern physics, such as condensed
matter, field theory, plasma physics, complex systems, dynamical
systems, and hydrodynamics~\cite{scaling1, scaling2}.
The \emph{universality} quality means that many different
systems present the same critical behaviour, while \emph{scaling},
is concerned with the fact that in a neighborhood of a critical
point the system is scale invariant. Preserving this
symmetry in the system makes it possible to relate physical phenomena
which take place at very different length scales. As a consequence, the
correct description of systems near their critical points can be
described by \emph{power laws} and this kind of behaviour might be
analysed by dimensional considerations known as \emph{scaling laws}.
Even though the Renormalization Group (RG) approach is a good alternative to obtain in
an accurate way the critical exponents~\cite{scaling7, scaling8}, in
many occasions the use of this approach in complex systems is
quite difficult. On the other hand, numerical simulations allow one to
describe in a simpler and more attractive way different complex systems,
but the possibility to use numerical simulations near the critical
points of a system is still a topic under discussion. \emph{Coarse
graining} is another common concept when we study systems which
present scale invariance and, when different scales are involved,
the coarse grained simulations have shown to be a very good
alternative. DPD simulations~\cite{scaling9} is one such coarse graining method and has shown that, if correct
parametrisations are used, can reproduce in great detail the
scaling properties of different kinds of real systems.

\begin{figure}
\begin{center}
\scalebox{0.75}{\includegraphics{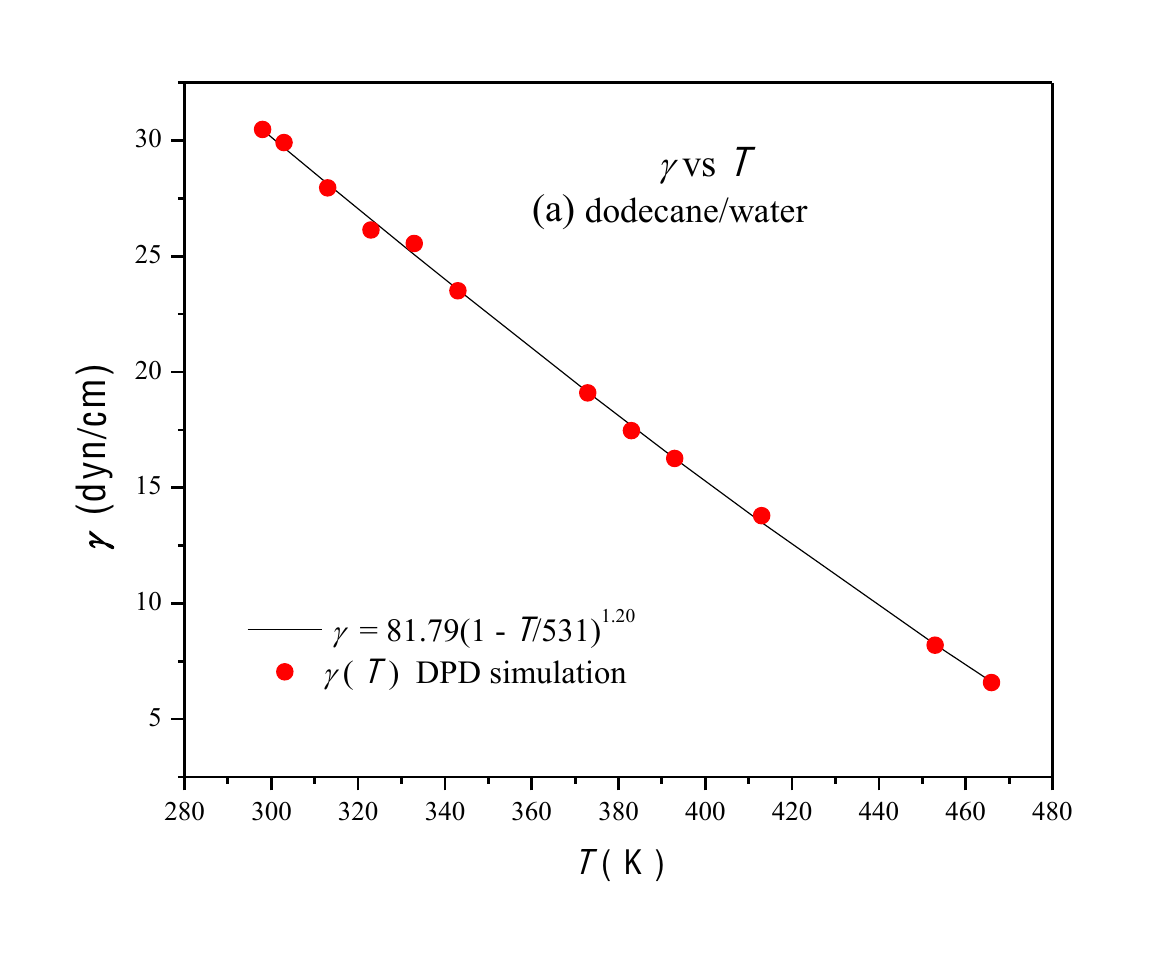}}
\caption{Scaling exponent observed for the dependence of the
interfacial tension $\gamma$ with temperature $T$ for dodecane/water}
\label{scalingTmu}
\end{center}
\end{figure}

As an example, the scaling exponent observed for the dependence of the
interfacial tension $\gamma$ with temperature $T$, for several liquid-liquid systems,
is given by:

\begin{equation}
   T = \gamma_o( 1-\frac{T}{T_c})^\mu
\end{equation}

\noindent where $\gamma_0$ is a system-dependent
constant, $T_c$ is the critical temperature
at which the interface becomes unstable, and $\mu$ is a critical exponent
which has been found experimentally some years ago to be close to
$11/9$~\cite{scaling17}. According to the
hyper-scaling relationship of Widom~\cite{scaling5, scaling6}, we have $\mu=\nu(d - 1)$ where $\nu$ is the scaling exponent for the radius of gyration given in eq.(~\ref{rgscaling}), and $d$ is the dimensionality of the system. More recently, by renormalization group
calculations~\cite{scaling7, scaling8, scaling18}, more accurate results
give us $\mu= 1.26$, and $\nu= 0.63$, which for $d=3$ satisfy the hyper-scaling law. These results have been reproduced by DPD simulations for different systems~\cite{mayoralSpringer1, mayoral5} and are presented in Figure~\ref{scalingTmu} for a dodecane/water mixture.

\begin{figure}
\begin{center}
\scalebox{0.60}{\includegraphics{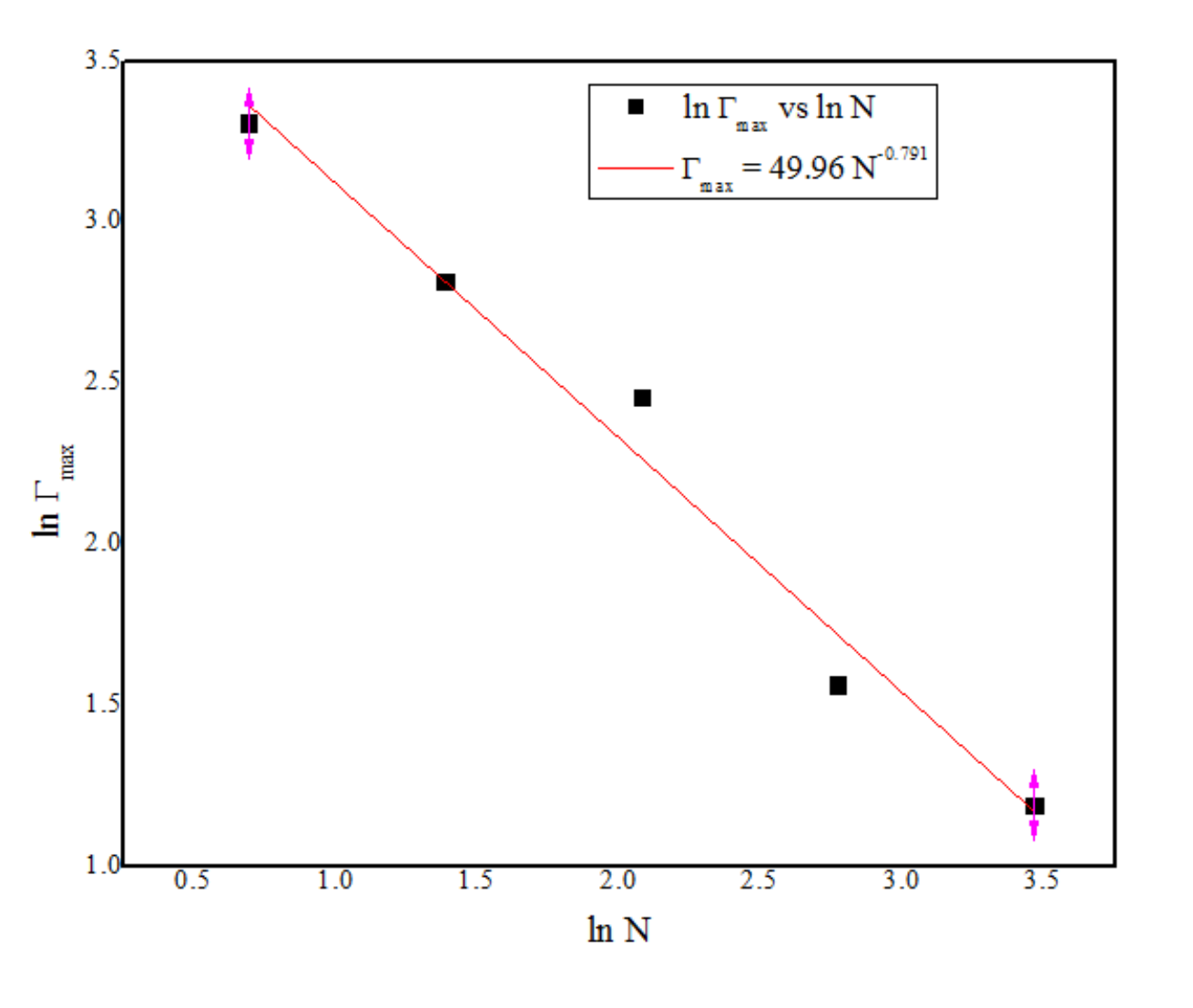}}
\caption{Scaling of $\gamma_{max}$ (maximum adsorption) with $N$ for polyacrylic acid on $TiO_{2}$ surfaces}
\label{scaling2}
\end{center}
\end{figure}

Another interesting example is the scaling of $\gamma_{max}$ (maximum adsorption) with the number $N$ of chain units.  The number of chains of size $N$ per unit area, $\Gamma_{max}$, needed to satisfactorily cover some given amount of material, say $1$ mol, can be obtained by performing DPD simulations for the adsorption of polymers with different length $N$ and fitting each simulation to a Langmuir isotherm. When  $\Gamma_{max}$ vs $N$ is plotted, the behaviour shown in Figure~\ref{scaling2} is obtained and the scaling function is $\Gamma_{max}\propto N^{-0.79}  \sim N^{-4/5}$. This result is in perfect agreement with the scaling theory in the weak adsorption regime~\cite{deGennes4}, which indicates that at maximum saturation

\begin{equation}
\gamma_{p} = \Gamma_{max} N \sim  N^{1/5}
\end{equation}

\noindent where $\gamma_{p}$ is the number of monomers adsorbed in the flat plateau of the isotherm. This implies $\Gamma_{max}\sim  N^{-4/5} = N^{-0.8}$ as obtained above.

Finally, another clear example are the scaling laws observed between the viscosity ($\eta$) and the friction coefficient ($\mu$). This behaviour was reproduced by non-equilibrium DPD simulations for sheared polymer chains grafted onto flat surfaces~\cite{mayoral6}. The scaling laws $\eta \sim \gamma^{-0.31}$ and $\mu \sim \gamma^{-0.69}$ at high shear rates $\gamma$ were obtained~\cite{mayoral6}.

\section*{Conclusions}

The appropriate parametrisation for the relevant parameters in Dissipative Particle Dynamics (DPD) simulations were presented. A clear methodology has been developed in the last few years to obtain the interaction parameters in great detail for realistic systems, making possible the study of their dependence with concentration and temperature. This work has proven to give predictions in accordance with experimental results. Explicit examples of interfacial tension, adsorption isotherms, disjoining pressure and radii of gyration are presented. Scaling properties present in different phenomena may also be reproduced in a precise manner using this methodology.

\section*{Acknowledgments}
This work was partially supported by DGAPA-UNAM (under project IN101614). Valuable support in computing resources was obtained from DGTIC-UNAM.

\end{document}